\documentclass[12pt,preprint]{aastex}

\shorttitle{Outflow - Core interaction in B1}
\shortauthors{Hiramatsu, Hirano, Takakuwa,  et al.}

\begin{document}


\title{Outflow -- Core Interaction in Barnard 1}


\author{Masaaki Hiramatsu\altaffilmark{1,2}, Naomi Hirano\altaffilmark{1}, and Shigehisa Takakuwa\altaffilmark{1}}


\altaffiltext{1}{Academia Sinica Institute of Astronomy and Astrophysics, P.O. Box 23-141, Taipei 10617, Taiwan; 
hiramatsu@asiaa.sinica.edu.tw}
\altaffiltext{2}{Institute of Astronomy, National Tsing Hua University, Hsinchu 33302, Taiwan}


\begin{abstract}
In order to study how outflows from protostars influence the physical and chemical conditions of the parent molecular 
cloud, we have observed Barnard 1 (B1) main core, which harbors four Class 0 and three Class I sources, in the 
CO ($J=1-0$), CH$_3$OH ($J_K=2_K-1_K$), and the SiO ($J=1-0$) lines using the Nobeyama 45 m telescope. We have identified three CO 
outflows in this region; one is an elongated ($\sim 0.3$ pc) bipolar outflow from a Class 0 protostar B1-c in the submillimeter clump
SMM 2, another is a rather compact ($\sim 0.1$ pc) outflow from a Class I protostar B1 IRS in the clump SMM 6, 
and the other is an extended outflow from a Class I protostar in SMM 11.
In the western lobe of the SMM 2 outflow, both the SiO and CH$_3$OH lines show broad redshifted wings 
with the terminal velocities of 25 km s$^{-1}$ and 13 km s$^{-1}$, respectively. It is likely that the shocks caused by 
the interaction between the outflow and ambient gas enhance the abundance of SiO and CH$_3$OH in the gas phase. 
The total energy input rate by the outflows ($1.1\times10^{-3} L_{\sun}$) is smaller than the energy loss 
rate ($8.5\times10^{-3} L_{\sun}$) through the turbulence decay in B1 main core, which suggests that the outflows can not sustain 
the turbulence in this region. Since the outflows are energetic enough to compensate the dissipating turbulence energy in the 
neighboring, more evolved star forming region NGC 1333, we suggest that the turbulence energy balance depends on the 
evolutionary state of the star formation in molecular clouds.
\end{abstract}


%% subject header.  Objects should be in the appropriate "individual"
%% headers (e.g. quasars: individual, stars: individual, etc.) with the
%% additional provision that the total number of headers, including each
%% individual object, not exceed six.  The \objectname{} macro, and its
%% alias \object{}, is used to mark each object.  The macro takes the object
%% name as its primary argument.  This name will appear in the paper
%% and serve as the link's anchor in the electronic edition if the name
%% is recognized by the data centers.  The macro also takes an optional
%% argument in parentheses in cases where the data center identification
%% differs from what is to be printed in the paper.



\section{INTRODUCTION}
Molecular outflows are common features accompanied with young stellar objects \citep[YSOs;][]{Bachiller1996}. 
The outflows have large impacts on the surrounding molecular cloud both in the chemical and physical aspect. 
Shocks excited by the propagation of the outflow raise the temperature of surrounding material and liberate 
various molecules such as NH$_3$, CH$_3$OH and SiO from dust grains 
\citep{Allamandola1992, Schilke1997, Bachiller2001, Jorgensen2004}. As a result, the relative abundance of these 
molecules against H$_2$ reaches several orders of magnitude larger than that in quiescent molecular clouds.

The supersonic turbulence in molecular clouds is ubiquitous but should quickly dissipate through shocks, thus some 
driving sources are needed \citep{MacLow1998}.
Protostellar outflows are considered to be possible sources to maintain the supersonic turbulence in molecular clouds 
\citep{Norman1980} and regulate the star formation, 
since the turbulence serves as a counterwork against the gravitational collapse in molecular clouds.
The observational studies of  Herbig-Haro (HH) objects \citep{Walawender2005a} and CO outflows 
\citep{Knee2000, Stanke2007} compared the energy input by the outflows and turbulence energy decay, and found that
the outflows have enough energy to retain the turbulence in the molecular clouds in at least local scale ($\sim 0.5$ pc), 
such as NGC 1333 or L1641-N, but not whole Perseus or Orion giant molecular cloud scale ($\ga 10$ pc).
The theoretical simulation also pointed out that the protostellar outflows play an important role to retain the turbulence and 
regulate the star formation efficiency \citep{Li2006}.

Barnard 1 (B1) is one of the star-forming regions located in the Perseus molecular cloud complex. 
The distance to B1 is estimated to be 230 pc by \citet{Cernis2003} on the basis of the extinction study. This value 
is consistent with the distance of 235 $\pm$ 15 pc to the NGC 1333 molecular cloud, which is located at $\sim 1\arcdeg$ west of B1,
derived from the parallax measurements of the H$_2$O masers associated with the YSO SVS 13 \citep{Hirota2008}. 
Note that this distance is different from the previously adopted values of 350 pc and 318 pc, which are the distances to the Perseus OB2 
association \citep{Borgman1964,deZeeuw1999}. 
B1 was observed in several molecular lines by \citet{Bachiller1990}. They revealed that the dense part of the cloud traced by 
the C$^{18}$O ($1-0$) line has a size of $\sim 1.3\times3.5$ pc and a mass of 518 $M_{\sun}$. Inside this region, a 
denser portion called ``main core'' with a diameter of 0.5 pc and a mass of 65 $M_{\sun}$ \footnote{These values are corrected for the
difference of the assumed distance from 350 pc to 230 pc.} was found in the NH$_3$ emission. A submillimeter continuum survey 
\citep{Walawender2005b} identified 12 submillimeter clumps in B1 and six of them are located in the main core region. Two of the 
submillimeter 
clumps (SMM 1 and SMM 2) harbor Class 0 YSOs and one (SMM 6) harbors a Class I YSO. Inside SMM 1, there are two extremely 
young sources, B1-bN and B1-bS \citep{Hirano1999}. A Class I source in SMM 6 was detected by \textit{IRAS} and is referred as
IRAS 03301+3057 or B1 IRS, and accompanied by an outflow \citep{Bachiller1990, Yamamoto1992, Hirano1997, Gregorio2005}. 
The YSO in SMM 2 (also referred as B1-c) located at $\sim 1\arcmin$ north of
IRAS 03301+3057 is also a driving source of the outflow that is extending along the east-west direction.
In addition to those four YSOs (B1 IRS, B1-bN, B1-bS, and B1-c), one Class 0 YSO was identified in SMM 3, and one Class I source 
in SMM 11 \citep{Hatchell2007a}. 
There is another Class I source, LkH$\alpha$ 327, without any submillimeter counterpart. In total, four Class 0 and three 
Class I sources are identified in B1 main core. The properties of the YSOs in B1 main core are summarized in Table \ref{sourceparam}.
Near infrared observations in H$\alpha$ and H$_2$ rovibrational lines found a number of shocks produced by protostellar outflows 
in the B1 main core region \citep {Walawender2005a, Walawender2005b, Walawender2009}. The most prominent shocks are associated with the outflow 
from SMM 2. Mid-infrared observations with \textit{Spitzer Space Telescope} have clearly revealed the S-shaped features of the 
knotty chains of the shocks centered at SMM 2, as well as several shocked regions around SMM 1, SMM 6, and SMM 11 
\citep{Jorgensen2006}.

This paper presents multi-molecular line observations of B1 main core to investigate the effects of protostellar outflows on the
physical and chemical properties of the surrounding core. The details of our observations are described in Section 2.
The results of the observations in CO, CH$_3$OH, and SiO lines are described in Section 3, followed by the
discussion on the chemical abundance of the molecules and turbulence energy budget in Section 4 and a summary 
in Section 5.

\section{OBSERVATIONS}

%% In a manner similar to \objectname authors can provide links to dataset
%% hosted at participating data centers via the \dataset{} command.  The
%% second curly bracket argument is printed in the text while the first
%% parentheses argument serves as the valid data set identifier.  Large
%% lists of data set are best provided in a table (see Table 3 for an example).
%% Valid data set identifiers should be obtained from the data center that
%% is currently hosting the data.

CO ($J=1-0$), CH$_3$OH ($J_K=2_K-1_K$), and SiO ($J=1-0$) line observations were performed in 2006 April with the 45 m 
telescope at Nobeyama Radio Observatory (NRO). The frequencies of the observed lines are summarized in Table \ref{lineparam}. 
The 25-Beam Array Receiver System \citep[BEARS;][]{Sunada2000} operated in the double sideband (DSB) mode was used 
for the CO and CH$_3$OH observations. The system noise temperature of BEARS during the observations was in the range of 
$T_{\rm sys}$ (DSB) = 200 -- 500 K. The main beam efficiency was $\eta _{\rm MB} \sim$ 0.45 at 115 GHz. The half power beamwidth 
(HPBW) was 18\arcsec at 97 GHz (CH$_3$OH) and 15\arcsec at 115 GHz (CO), and the separation between each beam was 41\farcs 1. Twelve 
pointings of BEARS were adopted to map the $10\arcmin \times 6\farcm5$ region, which covers the entire region of B1 main core, 
with a grid spacing of 29\farcs1. The intensity scales of the 25 receiver beams of BEARS at 97 GHz were calibrated by comparing the 
intensity of NGC 2264 observed by each beam of BEARS with that measured by the single-beam receiver S100. The intensity 
scales of the 25 beams at 115 GHz were calibrated 
using the scaling factor table provided by the observatory. The back end was an autocorrelator with a bandwidth of 32 MHz 
and a frequency resolution of 37.8 kHz, which corresponds to a velocity resolution of 0.098 km s$^{-1}$ and 0.12 km s$^{-1}$ 
at 115 GHz (CO) and 97 GHz (CH$_3$OH), respectively.

The SiO ($J=1-0$) line was observed along the jets from SMM 2 and SMM 6 with a single-beam SIS receiver S40 operated 
in the single-sideband (SSB) mode. The position angles of the strip scans are 77$\arcdeg$ and 133$\arcdeg$ for SMM 2 and SMM 6, 
respectively. The observational grid spacing was 38\farcs4, which was the same as the HPBW at 43 GHz. The
system noise temperature $T_{\rm sys}$ (SSB) during the observations was $\sim 180-300$ K. The main 
beam efficiency was $\eta _{\rm MB} = 0.77$ at 43 GHz. The bandwidth of the autocorrelator was set to 
32 MHz. The spectral resolution was 37.8 kHz which corresponds to a velocity resolution of 0.26 km s$^{-1}$ at 43 GHz. 

The telescope pointing was checked by observing the SiO maser emission from NML Tau every 60-90 minutes
and the error was estimated to be less than $\sim 6\arcsec$ in 100 GHz and 10$\arcsec$ in 43 GHz observations.
The standard chopper-wheel method \citep{Kutner1981} was adopted to convert the received signal into the antenna temperature.
The data were reduced using NEWSTAR, a data reduction software developed by NRO. Linear baselines were
subtracted in most cases.


%% This section also includes several of the displayed math environments
%% mentioned in the Author Guide.

\section{RESULTS}
\subsection{Molecular Outflow Traced by the CO Line}
\subsubsection{Distribution of the Outflows}
\begin{figure}
\epsscale{.70}
\plotone{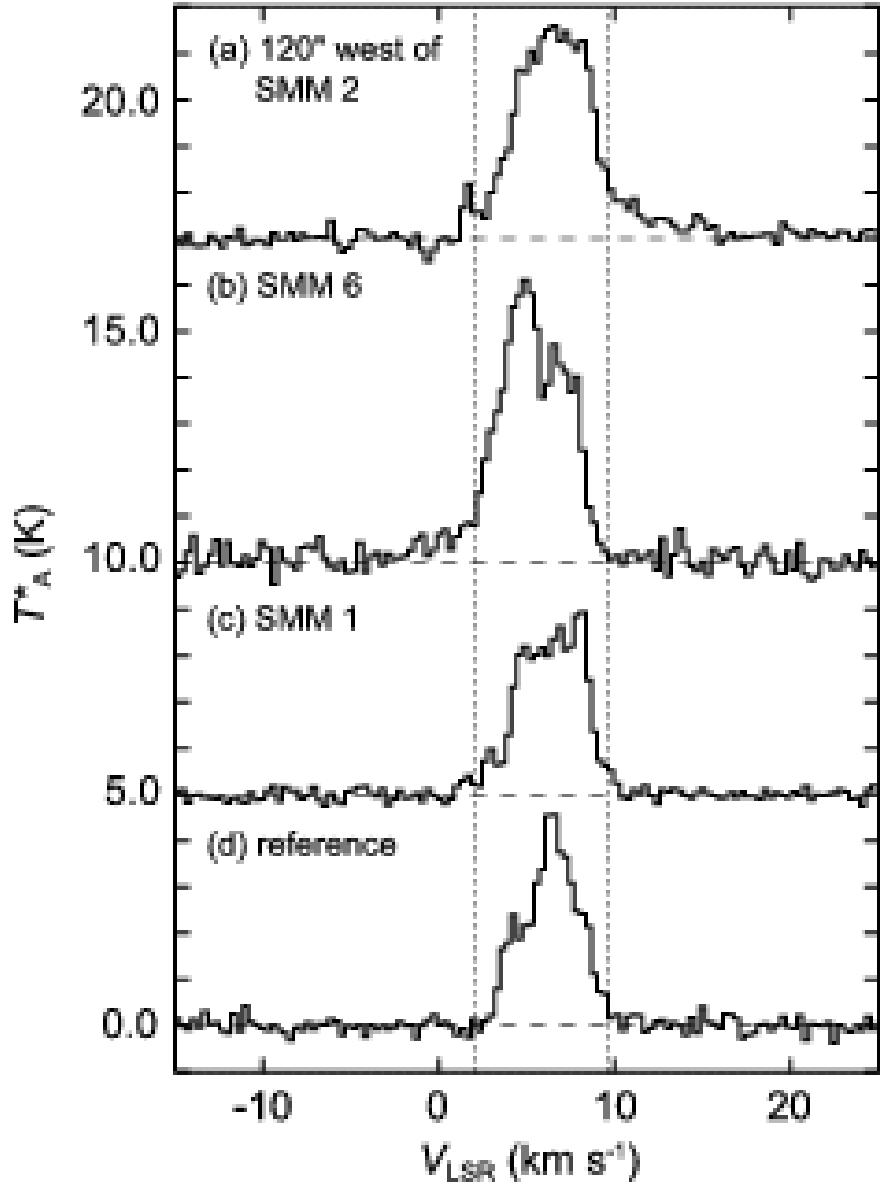}
\caption{CO ($J=1-0$) line profiles toward (a) $120\arcsec$ west of SMM 2, (b) SMM 6, (c) SMM 1 and (d) $5\farcm5$ southwest of SMM 6.
The velocity resolution is 0.3 km s$^{-1}$. Vertical lines indicate the outflow velocity boundaries.
\label{COlines}}
\end{figure}

Figure \ref{COlines} displays the CO ($1-0$) line profiles observed at four representative positions in B1 main core, i.e. 
(a) $120\arcsec$ west of SMM 2, (b) SMM 6, (c) SMM 1, and (d) 5\farcm5 southwest of SMM 6. In this paper, these four 
positions are referred to as positions ``a'', ``b'', ``c'', and ``d'', respectively. Since there is no jet-like component around this position
in the infrared image \citep{Walawender2005b}, it is likely that the spectrum observed at the position d is not affected by outflows.
Therefore, the velocity range of the ambient cloud component, ``line core'', was determined to be from $V_{\rm LSR}$ = 2.1 km 
s$^{-1}$ to $V_{\rm LSR}$ = 9.6 km s$^{-1}$, using the CO spectrum at the position ``d''. 
The CO emission outside of this velocity range is regarded as the outflow component. The CO spectra obtained at the positions
``a'' and ``b'' exhibit 
line wings in the redshifted side and blueshifted side, respectively. The most prominent CO redshifted wing was observed at the
position ``a''. 
At the position ``c'', wing-like blueshifted emission is detected in the velocity range of 1 km s$^{-1}$ $< V_{\rm LSR}< 3$ km
s$^{-1}$. This component might be originated from an outflow associated with SMM 1, however, because of its low velocity, it is 
unclear whether this component arises from the outflow or not. In addition, the spatial distribution of the blueshifted CO wing
(see below) is not localized on SMM 1. Therefore, in this paper we do not include this component as an outflow candidate.

\begin{figure}
\epsscale{.90}
\plotone{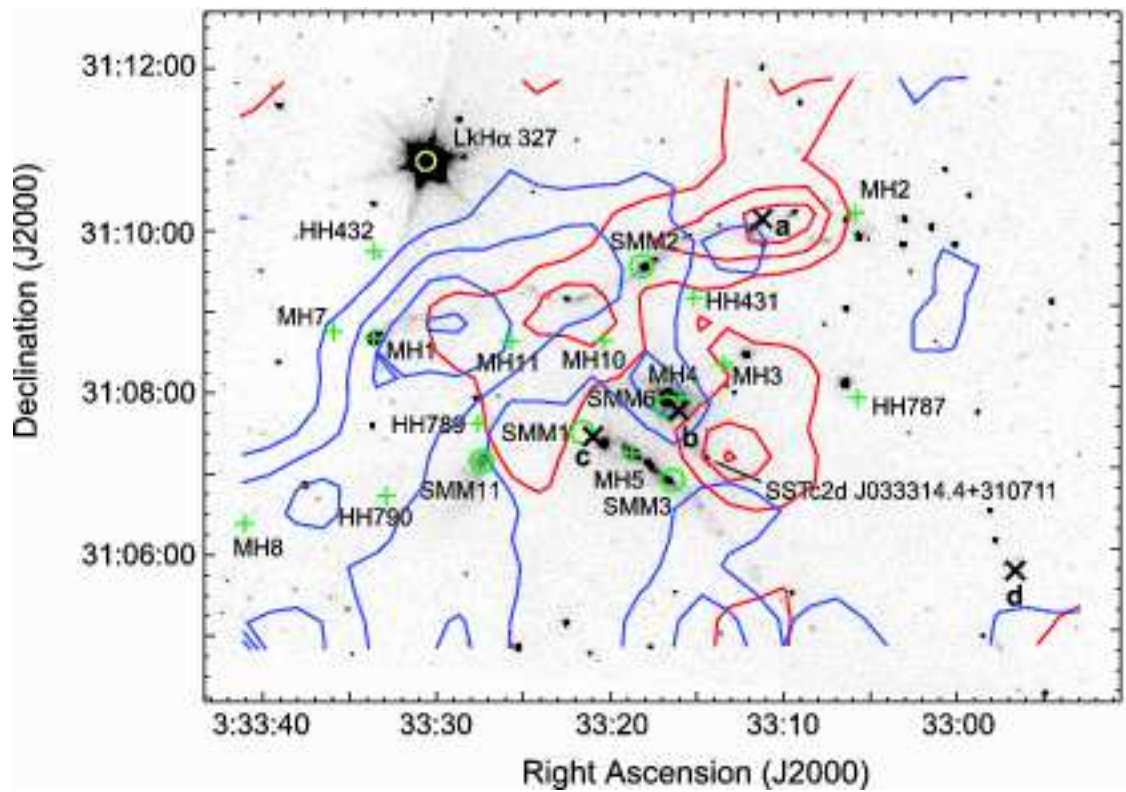}
\caption{Distribution of the high velocity component of the CO ($J=1-0$) line. The blueshifted ($-8.1$ km s$^{-1}$
$\le V_{\rm LSR} \le +2.1$ km s$^{-1}$) and the redshifted (9.6 km s$^{-1}$ $\le V_{\rm LSR} \le$ 19.8 km
s$^{-1}$) components are drawn with blue and red contours, respectively. 
The first contour level and the contour interval are 3$\sigma$, which correspond to 0.5 K km s$^{-1}$ for blue and
0.68 K km s$^{-1}$ for red contours, respectively. The gray-scale image in the background is the \textit{Spitzer} IRAC2
4.5 $\mu$m image. Circle and cross marks indicate the positions of protostars and H$_2$ knots \citep{Walawender2005b},
respectively. X marks indicate the positions ``a'' to ``d'' in Fig.\ref{COlines}.
 \label{COoutflow}}
\end{figure}

The spatial distributions of the blue- and redshifted outflow components are shown in Fig.\ref{COoutflow}. Although the spatial 
distributions of the high 
velocity components are complex, two bipolar outflows centered at SMM 2 and SMM 6, and the extended blueshifted emission at the southeast corner
are recognized.
The outflow centered at SMM 2 extends along the east$-$west direction. The extent of the westward
redshifted component is 120\arcsec (0.13 pc) and reaches the position of a shocked region MH 2 \citep{Walawender2005b}.
The redshifted emission peaks at the head of the lobe (position ``a'') and the velocity is also the largest at this position. 
On the other hand, the eastern blueshifted component peaks at 150\arcsec (0.17 pc)  southeast of SMM 2.
Neither the redshifted peak in the west nor the blueshifted peak in the east was covered by 
the previous CO ($3-2$) map \citep{Hatchell2007b} and the interferometric CO ($1-0$) map \citep{Matthews2006}.
Hence the entire picture of this outflow has been revealed by our observations
for the first time. There is a compact blueshifted component in the western lobe and an extended redshifted component
in the eastern lobe. This suggests that the outflow axis is close to the plane of the sky, and the overlapping blueshifted and redshifted
components are due to the projection of the front and back sides of the lobes. The outflow traced by the CO line is well aligned 
with the S-shaped knots seen in the \textit{Spitzer} IRAC band 2 image \citep{Jorgensen2006}. This IRAC band contains several H$_2$ 
pure-rotational lines that are considered to arise from shocked regions. Therefore, it is likely that the S-shaped feature 
traces the shocks in the outflow driven by SMM 2.

The outflow from SMM 6 is more compact than that from SMM 2. The blueshifted emission is detected toward the position
of SMM 6 and slightly extended ($\sim40\arcsec$, 0.04 pc) to the northeast. There is a redshifted component which peaks at 
$65\arcsec$ (0.07 pc) southwest of SMM 6. It is uncertain whether this redshifted component is originated from the SMM 6 
outflow or not, because there is another YSO 
candidate SSTc2d J033314.4+310711 at $20\arcsec$ (0.02 pc) east of the redshifted peak. This source is classified as a 
``YSOc\_red'' (YSO candidate and red) in the \textit{Spitzer} c2d catalog, and might be the driving source of the redshifted outflow. 
However, the grid spacing of the current observations
is not fine enough to pinpoint the position of the driving source unambiguously. Therefore, in the following discussion, the redshifted 
component at $65\arcsec$ southwest of SMM 6 is treated as a redshifted lobe of the SMM 6 outflow.
A series of shocked knots MH 4 was found at $15\arcsec$ west of SMM 6 and thought to be emanated from SMM 6
\citep{Walawender2005b, Walawender2009}. These knots align east--west direction and are not parallel to the SMM 6 CO outflow.
Since the length of the knots is $20\arcsec$, smaller than our mapping grid spacing, the relation between the knots and the SMM 6 
outflow is not clear in the present observations. An elongated reflection nebula was identified in the $K_s$-band image around SMM 6
\citep{Walawender2005b} and the orientation is consistent with that of the SMM 6 outflow. This consistency would support the
idea that SMM 6 is the driving source of the redshifted component southwest of SMM 6.

The extended blueshifted CO emission is found around the shocked knots MH 8 and HH 790. 
In the CO image, this blueshifted component is continuously connected to the blueshifted lobe of the SMM 2 outflow. If this blueshifted  
component in the southeast area is a part of the SMM 2 outflow, however, the outflow is highly asymmetric with a blueshifted lobe more than twice  
as longer as the redshifted lobe. In addition, the blueshifted lobe bends sharply to the south at the position around the MH 1 knot. In  
contrast, the shocked components seen in the \textit{Spitzer} image suggest that the eastern lobe and the western lobe of the SMM 2 outflow are similar  
in length. Therefore, it is possible that the blueshifted emission in the southeastern area originates from the other outflow. One possible  
candidate is the giant east--west outflow from SMM 11, which was identified in the deep H$_2$ image obtained by \citet{Walawender2009}.  
This outflow is parallel to the SMM 2 outflow, and contains MH 8 and HH 790 in the eastern lobe. If MH 8 and HH 790 are related to the outflow  
from SMM 11, it is natural to consider that the blueshifted CO component around MH 8 and HH 790 is also a part of the SMM 11 outflow. 
In the H$_2$ image, the western lobe of the SMM 11 outflow extends to MH 10 and MH 3. The redshifted CO emission was detected around MH 10 
and would be the part of the SMM 11 outflow. However, the giant outflow extending further west was not clearly detected in our CO observations.

The outflows from SMM 1, SMM 3, and LkH$\alpha$ 327 are not clearly seen in our CO ($1-0$) map, although they were
reported by \citet{Hatchell2007b} on the basis of the CO ($3-2$) observations, except for that from LkH$\alpha$ 327. 
Our CO ($1-0$) maps show the redshifted emission to the northeast of SMM 1. This component is also seen in the CO ($3-2$) map of 
\citet{Hatchell2007b}, and identified as the redshifted lobe of the SMM 1 outflow. It is possible that this redshifted 
emission is related to the outflow from SMM 1 as argued by \citet{Hatchell2007b}. However, we regard this emission as a part
of the outflow from SMM 11 because there is no clear blueshifted component and bipolarity around SMM 1.
The mid- and near infrared images show a chain of shocked knots that extends from SMM 1 to SMM 3. The axis of this knot 
complex is parallel to that of the SMM 6 outflow. Although SMM 1 is designated to be 
a driving source of these shocked features \citep{Walawender2009}, no corresponding CO high-velocity component was detected 
in the present observations. 
LkH$\alpha$ 327 was classified as Class I  based on the spectral energy distribution 
slope in the  \textit{IRAS} $12-100$ $\micron$ data \citep{Walawender2005b}. On the other hand, this source was classified as Class II 
based on the Two Micron All Sky Survey \textit{K}-band, \textit{Spitzer} IRAC, and MIPS data. The facts that LkH$\alpha$ 327 is located at 
the periphery of B1 main core and not associated with any 850$\micron$ emission indicate that the source is in the rather 
evolved stage. 

\subsubsection{Estimation of the Kinematic Properties of the Outflows}
The mass, momentum, kinetic energy, and the momentum flux of the identified CO outflows are calculated with the method 
described by \citet{Bourke1997}. In this procedure, we divided the high-velocity CO emissions originated from the outflows into the 
0.4 km s$^{-1}$ velocity bins, and 
calculated the quantities on an assumption of the local thermal equilibrium (LTE) condition. First, the column density of
the molecule in the velocity range of $v$ to $v+\Delta v$, $N(v)$, is given as
\begin{eqnarray}
N(v)&=&{3hZ(T) \over 8\pi^3\mu^2(J+1)}{\exp\{hBJ(J+1)/kT_{\rm ex}\} \over 1-\exp(-h\nu/kT_{\rm ex})}f\int_v^{v+\Delta v} \tau dv, 
\end{eqnarray}
where
\begin{eqnarray}
f &=& {T_{\rm MB} \over J(T_{\rm ex})-J(T_{\rm bg})}{1 \over 1-\exp(-\tau)},\\
J(T)&=&{h\nu/k \over \exp(h\nu/kT)-1},
	\label{cd}
\end{eqnarray}
$h$ is the Planck constant, $Z(T)$ is the partition function, $\mu$ is the dipole moment of the molecule,
$B$ is the rotation constant of the molecule, $k$ is the Boltzmann constant, $\nu$ is the frequency of the line, 
$\tau$ is the optical depth, $T_{\rm ex}$ is the excitation temperature, and $T_{\rm bg}$ is the background temperature. 
For linear molecules such as CO, the partition function is approximated to be $Z(T)=kT/hB$ if $hB \ll kT$.
In the case of the CO ($1-0$) line under the optically thin assumption, the above equations become the following as given by 
\citet{Bourke1997}:
\begin{eqnarray}
N(v) = 2.31 \times 10^{14} {{T_{\rm ex} + 0.92}\over{1-{\rm exp}(-5.53/T_{\rm ex})}} {1\over{J(T_{\rm ex})-J(T_{\rm bg})}}
\int_v^{v+\Delta v} T_{\rm MB} dv.
\end{eqnarray}
The mass at each observed position ($\alpha, \delta$) and velocity $v$, $M(v, \alpha, \delta)$ is given as	
\begin{eqnarray}
	M(v, \alpha, \delta) = [{\rm H}_2/{\rm CO}] \mu_{\rm m}m_{\rm H}A(\alpha, \delta)N(v),
\end{eqnarray}
where [H$_2$/CO] is the reciprocal of the CO abundance and adopted to be $10^4$, $\mu_{\rm m}=2.72$ is the mean molecular 
mass per H$_2$, $m_{\rm H}$ is the mass of a hydrogen atom, and $A(\alpha, \delta)$ is the area of the CO emission at a given 
observed position. Since the observing grid
spacing ($29\farcs1$) is larger than the telescope beam (15$\arcsec$), we adopted the squared value of the observing grid 
spacing as $A(\alpha, \delta)$ under the assumption of the uniform emission distribution in the $29\farcs1^2$ region.
The mass at each velocity bin $M(v)$ is estimated by integrating $M(v, \alpha, \delta)$ over the outflow spatial 
extension, and then the total mass is derived by integrating $M(v)$ over the outflow velocity range. 

Since the optical depth of the CO line is not obtained in our observations, the assumption of the optically thin CO ($1-0$)
line is adopted, hence the estimation should be a lower limit. The excitation temperature $T_{\rm ex}$ is
estimated by comparing the CO ($1-0$) spectra and ($3-2$) spectra \citep{Hatchell2007b} as follows. Under the LTE condition, the
total column density $N$ and the column density at a given rotational energy level $J$, $N_J$, is given as
\begin{eqnarray}
N=N_J{Z \over 2J+1}\exp\{hBJ(J+1)/kT_{\rm ex}\}.
\label{NtotLTE}
\end{eqnarray}
$N_J$ is also expressed as
\begin{eqnarray}
N_J={8\pi\nu^3 \over c^3}{2J+1 \over (2J+3)A_{J+1, J}}{1\over 1-\exp(-h\nu/kT_{\rm ex})} \int T_{\rm MB} dv
\label{NjTmb}
\end{eqnarray}
in the optically thin condition, where $A_{J+1, J}$ is the Einstein $A$ coefficient for the transition from the level $J+1$ to $J$.
The relation between the excitation temperature $T_{\rm ex}$ and the line intensity ratio of the CO ($3-2$) and ($1-0$) transition is 
derived from Equations (\ref{NtotLTE}) and (\ref{NjTmb}):
\begin{eqnarray}
{\int T_{\rm MB, 3-2} dv \over \int T_{\rm MB, 1-0} dv} = 3 \exp(-16.6/T_{\rm ex}) {1-\exp(-16.6/T_{\rm ex}) \over 1-\exp(-5.53/T_{\rm ex})}
 \label{lineratio}
\end{eqnarray}
The integrated intensities of the high-velocity CO ($3-2$) line toward the SMM 2 westward redshifted component and SMM 6 were 
calculated from the spectra shown by \citet{Hatchell2007b}. The integrated intensities, line ratios and the derived excitation 
temperatures are summarized in Table \ref{COlineratio}. Although the observed positions in the two transitions are slightly different, 
the offset ($< 5\arcsec$) is smaller than 1/3 of the beam size and we ignore the positional difference. The derived excitation 
temperatures toward the two positions, 46 K and 42 K, are close to each other. We note that the derivation of the excitation 
temperature is sensitive to the accuracy of the relative calibration between the two independent observations with the different
instruments. For example, a 20\% uncertainty of the ratio yields $\pm$ 20 K uncertainty of the excitation temperature.
Here we adopt 50 K as a value of the excitation temperature for the calculations of the outflow parameters. 
This value is the same as that used by \citet{Hatchell2007b} to calculate the outflow properties in the Perseus molecular cloud. 

The momentum $P$, kinetic energy $E_{\rm k}$, dynamical timescale $t_{\rm d}$, and the outflow momentum flux $F_{\rm outflow}$ 
are described as \begin{eqnarray}
	P &=& \int M(v)(v-v_0) dv,\\
	E_{\rm k}&=&{1\over 2}\int M(v)(v-v_0)^2 dv, \\
	t_{\rm d} &=& R / v_{\rm ch} = R / (P/M),\\
	F_{\rm outflow}&=&P/t_{\rm d},
\end{eqnarray}
where $v_0$ is the systemic velocity of 6.7, 6.5, and 6.8 km s$^{-1}$ for the SMM 2, SMM 6, and SMM 11 outflows, respectively, measured by fitting a 
Gaussian function to the CH$_3$OH spectra toward the sources (see Section 3.2). $R$ is the length of the outflow and 
$v_{\rm ch}=P/M$ is the characteristic velocity. The mass-loss rate is estimated as $\dot{M} = F_{\rm outflow} / v_{\rm wind}$, 
where $v_{\rm wind}$ is the velocity of the primary wind ejected from the close vicinity of the protostar. Since we do not have 
any information on the wind velocity from our observations, we assume $v_{\rm wind}=100$ km s$^{-1}$ \citep[e.g.,][]{Giovanardi1992}. 

The inclination angle of the outflow axis is also unclear, therefore we employed $57\fdg3$ from the line of
sight, which is the mean inclination angle on the assumption of the random outflow orientation.
This assumption should be reasonable for the SMM 2 outflow because the interferometric observations revealed that the opening angle of the outflow 
cavity in the 20\arcsec scale is $\sim 55\arcdeg$ and the red and blueshifted lobes are not overlapped at the position of the driving source 
\citep{Matthews2006}, which indicates that the outflow axis inclines at the larger angle than the opening angle. This interpretation
does not conflict with the fact that the blue- and redshifted components are overlapped in the larger scale as shown in 
Fig.\ref{COoutflow}, because these overlapped blue- and redshifted components are interpreted as the cavity wall. For the 
SMM 6 outflow, \citet{Hirano1997} suggested that the configuration of the SMM 6 outflow is in a pole-on geometry, while 
\citet{Gregorio2005} suggested that the outflow from SMM 6 is inclined toward southwest. Since it is difficult to estimate 
the inclination angle from our coarse grid observations and there is no consensus on the inclination angle of the SMM 6 outflow, 
it is reasonable to adopt the mean inclination angle. 
Since there is no previous study about the inclination angle for the SMM 11 outflow, we also adopt the mean inclination angle
of $57\fdg3$.

The blueshifted lobes of the outflows from SMM 2 and SMM 11 are not clearly separated in our observations. Therefore, we assumed that the  
transverse extent of the SMM 2 outflow lobe is symmetric with respect to the shocked knots, and regarded the blueshifted component at 
decl. $> 31\arcdeg 08\arcmin$ as the SMM 2 outflow and that at decl. $< 31\arcdeg 08\arcmin$ as the SMM 11 outflow. The redshifted lobe of the 
SMM 11 outflow is also confused with the eastern redshifted component of the SMM 2 outflow. We assumed that the northeastern half of the 
redshifted component at the east of SMM2 belongs to the SMM 2 outflow, because this part is overlapped with the blueshifted lobe of the SMM 2 
outflow. The southwestern half of this redshifted component is regarded as the redshifted lobe of the SMM 11 outflow.
The outflow parameters derived on the above assumptions are summarized in Table \ref{outflowparam}.

The mass, momentum, and kinetic energy of the SMM 2 outflow are almost comparable to those of the SMM 11 outflow
and a factor of 5--10 larger than those of the SMM 6 outflow.
The properties of outflows from Class 0 and Class I sources were compiled by \citet{Bontemps1996}, and it was found that
the outflow momentum flux is 1 order of magnitude larger for Class 0 sources. The mean momentum
flux is $<F_{\rm CO}>=5.7\times10^{-5} M_{\odot}$ km s$^{-1}$ yr$^{-1}$ and $3.8\times10^{-6} M_{\odot}$ 
km s$^{-1}$ yr$^{-1}$ for Class 0 and Class I sources, respectively \citep{Bontemps1996}. Considering that they
adopted an opacity correction of a factor of 3.5, the momentum fluxes of the outflows emanated from SMM 2 and SMM 6 are
in reasonable agreement with the mean values for Class 0 and I sources, respectively. 
On the other hand, the outflow from SMM 11 has approximately 1 order of magnitude larger momentum flux compared to
the mean value of the outflows from the Class I sources, probably because the spatial extent of this outflow is large.

The momentum fluxes for the outflows were estimated by \citet{Hatchell2007b} based on their CO ($3-2$) 
observations. Converting the assumed distance to B1 from 320 pc to 230 pc, the momentum fluxes are
$F_{\rm CO}=1.2\times10^{-7}$, $9.1\times10^{-7}$, 
and $3.2\times10^{-7} M_{\sun}$ km s$^{-1}$ yr$^{-1}$ for the SMM 2, SMM 6, and SMM 11 outflow, respectively. 
The momentum flux they derived for all the outflows is smaller than our result, and the difference reaches a factor of 70
for the SMM 2 outflow.
There are several reasons for these discrepancies. First, 
they did not take the correction of the inclination $i$ into account, and this correction makes the momentum flux larger by 
a factor of $\sin i/\cos^2i$ \citep{Bontemps1996}. Our assumption of the inclination of $57\fdg3$ yields the factor of 2.9. 
Second, we regard both the westward and eastward redshifted components as the outflow from SMM 2, however, 
\citet{Hatchell2007b} assigned the eastward redshifted component to the SMM 1 (B1-bS) outflow. This different identification 
causes another factor of 2 discrepancy in the estimation of the outflow momentum flux. Finally, the mapped region by 
\citet{Hatchell2007b} does not cover the entire extent of the outflow from SMM 2. In general, the outflow momentum flux is a robust 
parameter against incomplete mappings. If a considerable fraction of the total momentum is carried by components 
outside the mapping area, however, incomplete mapping observations might result in underestimation of the momentum flux. 
In fact, the prominent redshifted component around the position ``a'' is located outside of the mapped area by \citet{Hatchell2007b}. 

\subsection{CH$_3$OH}
\begin{figure}
\epsscale{.70}
\plotone{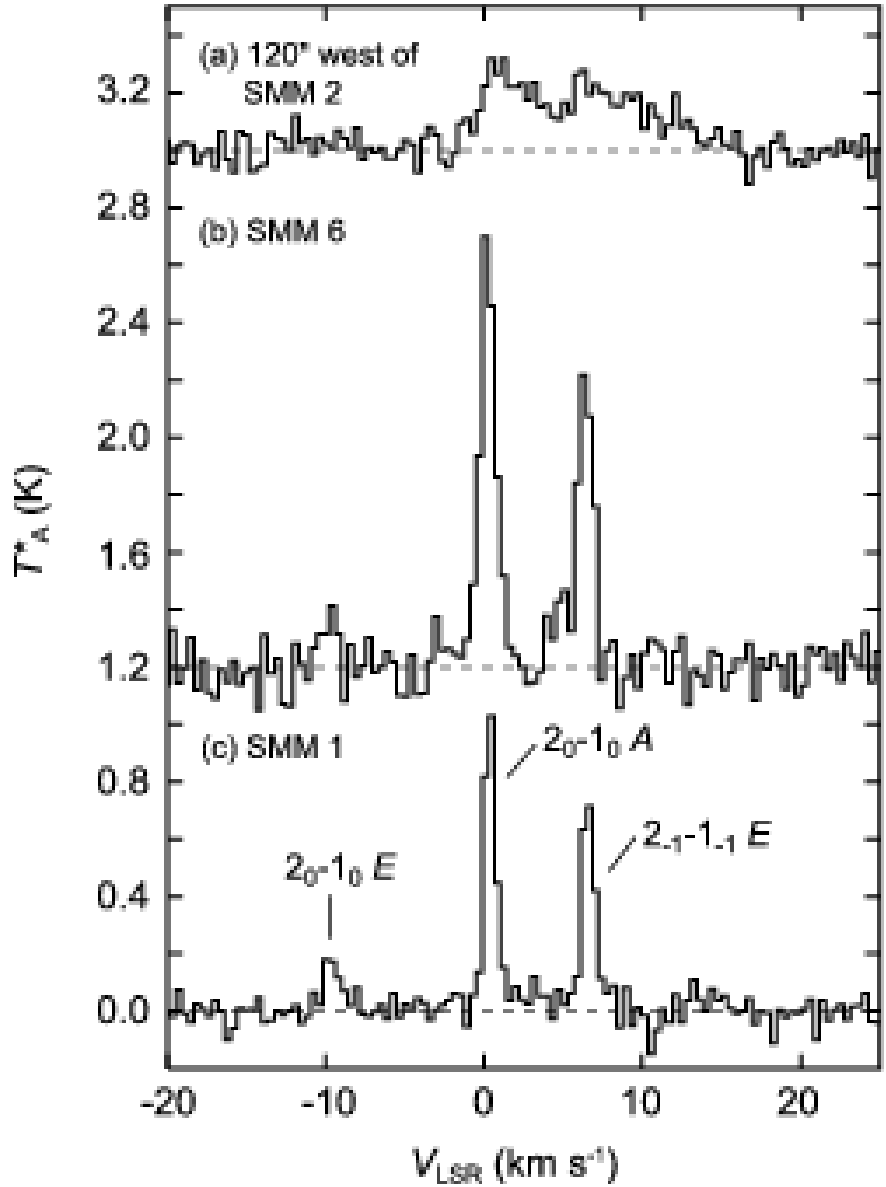}
\caption{CH$_3$OH ($J_K=2_0-1_0$ $E$, $=2_0-1_0$ $A$, and $2_{-1}-1_{-1}$ $E$) line profiles
toward (a) $120\arcsec$ west of SMM 2, (b) SMM 6, and (c) SMM 1.
The velocity resolution is 0.3 km s$^{-1}$. The reference frequency in the velocity calculation is set to 
be 96.73939 GHz, the rest frequency of $2_{-1}-1_{-1}$ $E$ transition.
\label{METlines}}
\end{figure}

The CH$_3$OH line profiles at three representative positions, i.e. 
(a) $120\arcsec$ west of SMM 2, (b) SMM 6, and (c) SMM 1, are displayed in Fig. \ref{METlines}.  
The frequency separations between the $2_0-1_0$ $A^+$ and the $2_{-1}-1_{-1}$ $E$ lines and the $2_0-1_0$ $E$ and the
$2_0-1_0$ $A^+$ lines correspond to 9.8 km s$^{-1}$ and 6.3 km s$^{-1}$, respectively.
The widest line profile is seen at the position ``a'' (Fig.\ref{METlines}\textit{a}) where the peak of the CO redshifted outflow is located. 
At this position, the $2_0-1_0$ $A^+$ and $2_{-1}-1_{-1}$ $E$ lines are blended because of 
the broad redshifted wings. The terminal velocity of the $2_{-1}-1_{-1}$ $E$ line reaches 13 km s$^{-1}$ away 
from the systemic velocity. The position ``a'' is located close 
($\sim 10\arcsec$) to the branching point of the shocked feature seen in the \textit{Spitzer} image (see Fig.\ref{COoutflow}). 
Similar blended 
CH$_3$OH profiles are seen in three observed points around the position ``a''. Toward SMM 6, both the $2_0-1_0$ $A^+$ and $2_{-1}-1_{-1}$ $E$
lines show a blueshifted wing (Fig.\ref{METlines}\textit{b}) which is also seen in the CO line (Fig.\ref{COlines}\textit{b}). No wing 
components are detected toward SMM 1(Fig.\ref{METlines}\textit{c}) where no CO outflow was identified. From these results
we divide the CH$_3$OH emission into three velocity ranges; the blueshifted component ($V_{\rm LSR} < 5.5$ km s$^{-1}$),
the line core component  ($5.5$ km s$^{-1} < V_{\rm LSR} < 7.5$ km s$^{-1}$), and the redshifted component ($V_{\rm LSR} > 7.5$ 
km s$^{-1}$).

\begin{figure}
\epsscale{1.0}
\plotone{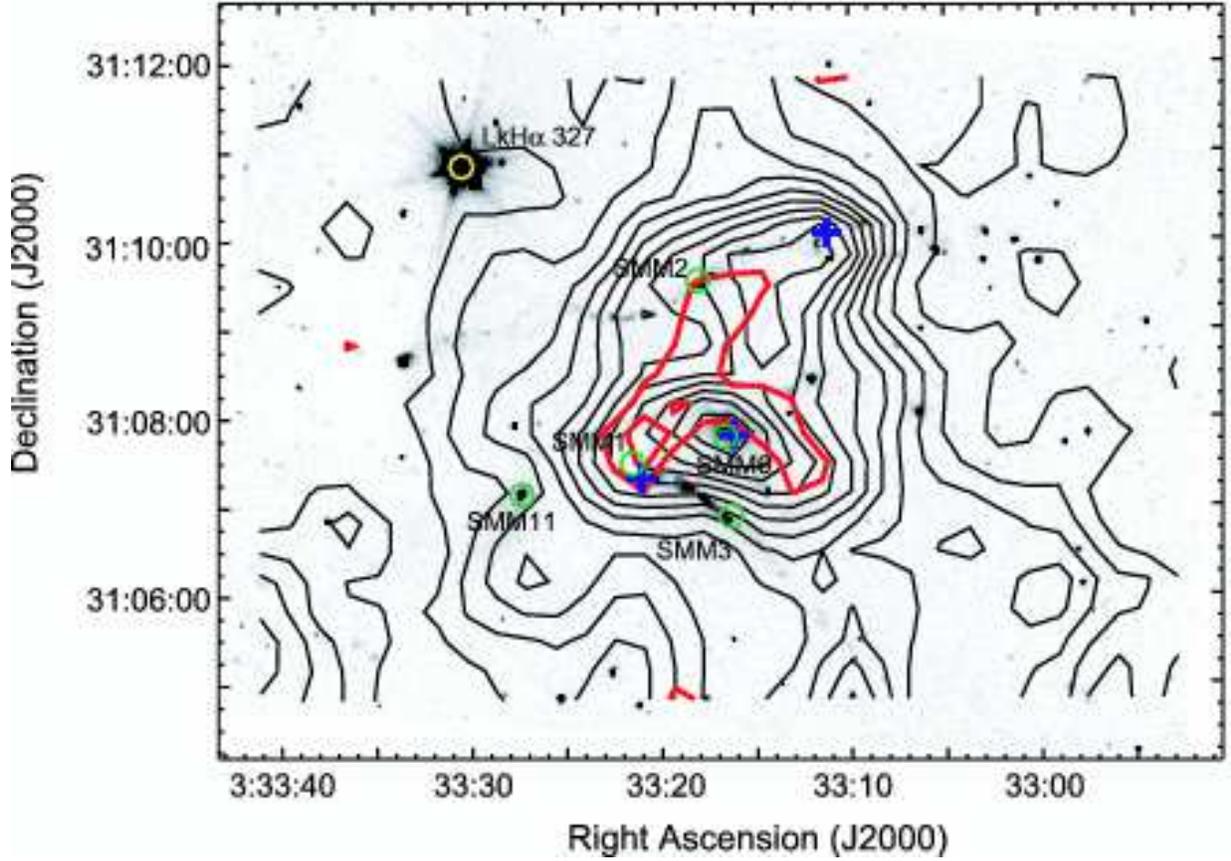}
\caption{Integrated intensity map of the CH$_3$OH lines. Black contours represent the integrated intensity of  the $J_K=2_{-1}-1_{-1}$ $E$ 
and $2_0-1_0$ $A^+$ lines ($A+E$ map). Red contours show the integrated intensity of the $2_0-1_0$ $E$ line. Both the lowest contour 
level and the interval are 0.23 K km s$^{-1}$ (6$\sigma$) for the $A+E$ map. The lowest contour level is 0.13 K km s$^{-1}$ and the contour 
interval is 0.043 K km s$^{-1}$ (1$\sigma$ noise level) for the $2_0-1_0$ $E$ line.  The integrated velocity range is $-3.6 < V_{\rm LSR} <$ 
20.4 km s$^{-1}$ for the $A+E$ map and 2.9 $< V_{\rm LSR} <$ 10.2 km s$^{-1}$ for the $2_0-1_0$ $E$ line. The position ``a'' and
the protostars are marked with a cross and circles. 
\label{METmap}}
\end{figure}

\begin{figure}
\epsscale{1.0}
\plotone{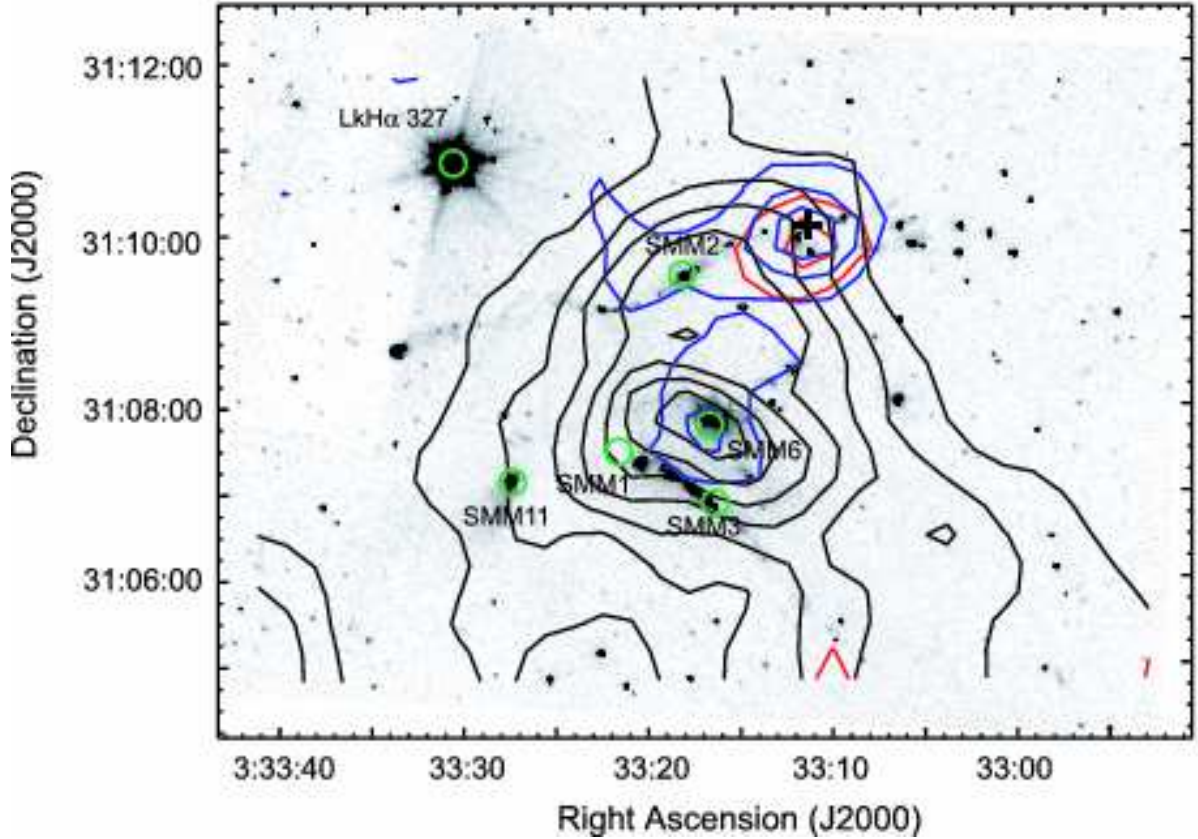}
\caption{Integrated intensity map of the CH$_3$OH ($J_K=2_{-1}-1_{-1}$ $E$) line in the line core velocity range (5.5 km s$^{-1}$ 
$< V_{\rm LSR} <$ 7.5 km s$^{-1}$, black contours) and the higher velocity range (2.5 km s$^{-1}$ $< V_{\rm LSR} <$ 5.5
km s$^{-1}$ in blue contours and 7.5 km s$^{-1}$ $< V_{\rm LSR} <$ 20 km s$^{-1}$ in red contours). Note that the
blueshifted component around the position ``a'' (black cross) is due to a contamination of redshifted emission of the $J_K=2_{0}-1_{0}$ $A^+$
line. The lowest contour and contour intervals are 3$\sigma$ noise level of each map, 0.12 K km s$^{-1}$ for the
line core, 0.09 K km s$^{-1}$ for the blueshifted, and 0.24 K km s$^{-1}$ for the redshifted component.
\label{METlcmap}}
\end{figure}

The integrated intensity maps of the CH$_3$OH lines are shown in Fig.\ref{METmap}. Since the distribution of the $2_0-1_0$
$A^+$ line is similar to that of the $2_{-1}-1_{-1}$ $E$ line, and these two lines are blended around the position ``a'', we integrated all
the $2_0-1_0$ $A^+$ and $2_{-1}-1_{-1}$ $E$ emission over the velocity of  $-3.6$ km s$^{-1} < V_{\rm LSR} < 20.4$ km s$^{-1}$
 with the reference frequency of  96.73939 GHz, the rest frequency of $2_{-1}-1_{-1}$ $E$ line, and made a combined map
which is shown in Fig.\ref{METmap}. Hereafter we call this map ``$A+E$ map''. 
The CH$_3$OH ($J_K=2_0-1_0$ $A^+$, $2_{-1}-1_{-1}$ $E$) lines were detected around the dusty clump complex which 
contains SMM 1, SMM 2, SMM 3, and SMM 6. There are two peaks seen in the $A+E$ map; 
one peak is located at the position ``a'' and the other peak is located around SMM 6. No obvious 
peaks are detected in the eastern side of the SMM 2 outflow and around the other protostars.  
The observed CH$_3$OH distribution is different from those of the dust continuum and NH$_3$ emission
\citep{Matthews2002, Walawender2005b, Bachiller1990}. Since one CH$_3$OH peak located at 
the position ``a'' coincides with the peak of the redshifted CO emission, we examined the contribution of the high-velocity  
component to the $A+E$ map by comparing the spatial distributions of the blue- and redshifted CH$_3$OH emission with 
that of the line core emission (Fig. \ref{METlcmap}). The high-velocity components peak at the position ``a'' and SMM 6, and it is
obvious that the $A+E$ map is largely affected by the high-velocity components. The line core emission shows a single peak at SMM 6,
and its distribution is different from the distribution of the NH$_3$ emission which peaks around SMM 1. The distribution of the 
CH$_3$OH line core emission is also different from the distribution of the dust continuum emission which peaks at the positions of
the SMM sources \citep{Walawender2005b}. The coincidence of the peaks of the line core component and the high velocity 
(blueshifted) component at SMM 6 indicates that the peak around SMM 6 seen in the $A+E$ map arises from both the quiescent 
gas and the outflow-related component. The CH$_3$OH abundance enhancement due to the outflow -- core interaction is a 
possible reason for the intensity enhancement around the outflows. The abundance enhancement is discussed in Section 4.1.

The CH$_3$OH $2_0-1_0$ $E$ transition is much weaker than the other two transitions and detected above the 3$\sigma$ noise level 
at eight out of the 300 observed positions. An example of the spectrum is shown in Fig.\ref{METlines}\textit{c}, which is 
detected at ($\alpha, \delta$)(J2000)=($3^{\rm h} 33^{\rm m}20 \fs8, 31\arcdeg 07\arcmin 25\farcs1$), $10\arcsec$ southwest 
of the Class 0 protostar SMM 1 (B1-bS). The integrated intensity map of the $2_0-1_0$ $E$ transition is shown in
Fig.\ref{METmap} with red contours. This line shows its intensity peak around SMM 1 and extension to SMM 2 and 6. 
The CH$_3$OH $2_0-1_0$ $E$ emission peak at SMM 1 is not seen in the other two transitions, whereas the high density tracers 
such as the NH$_3$ (1,1), H$^{13}$CO$^+$ ($J=1-0$) lines, and 850 $\mu$m continuum emission show the emission peaks around SMM 1 
\citep{Bachiller1990, Hirano1999, Matthews2002}. The upper state energy of the 
$2_0-1_0$ $E$ transition is 12.2 K, which is slightly higher than those of the other two transitions, 4.6 and 7.0 K for the $2_{-1}-1_{-1}$ $E$ 
and $2_0-1_0$ $A^+$ transition, respectively \citep[obtained from CDMS catalog, ][]{Muller2001}. Since CH$_3$OH emission tends to
be sub-thermally excited in star-forming regions \citep{Buckle2002}, the difference of the distributions between the $2_0-1_0$ $E$
transition and the other lower-excitation transitions is likely to reflect the density structure of B1 main core, as the $2_0-1_0$ $E$
line and the high-density tracers show the common peak at SMM 1.

\subsection{SiO Emission along the Outflows}
\begin{figure}
\epsscale{.80}
\plotone{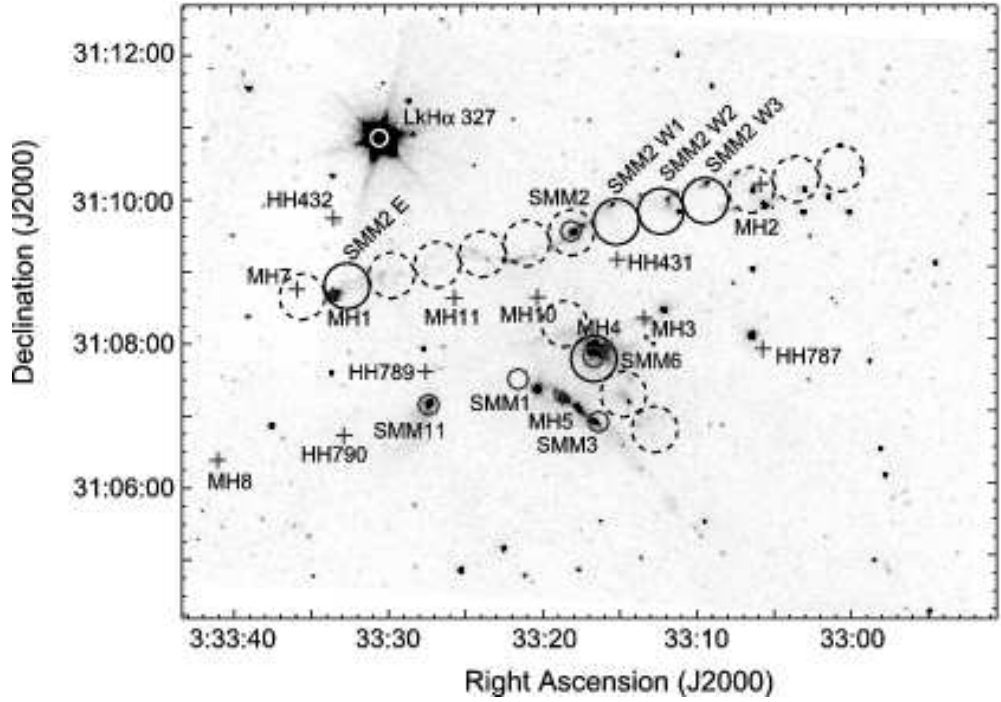}
\caption{SiO ($J=1-0$) observing positions overlaid on the \textit{Spitzer} IRAC2 image. The large circles with solid and 
broken lines indicate the positions where the SiO line was detected and not detected, respectively. The size of the 
circles corresponds to the HPBW in the frequency of the line (43.4 GHz). 
\label{SiOpos}}
\end{figure}

\begin{figure}
\epsscale{.80}
\plotone{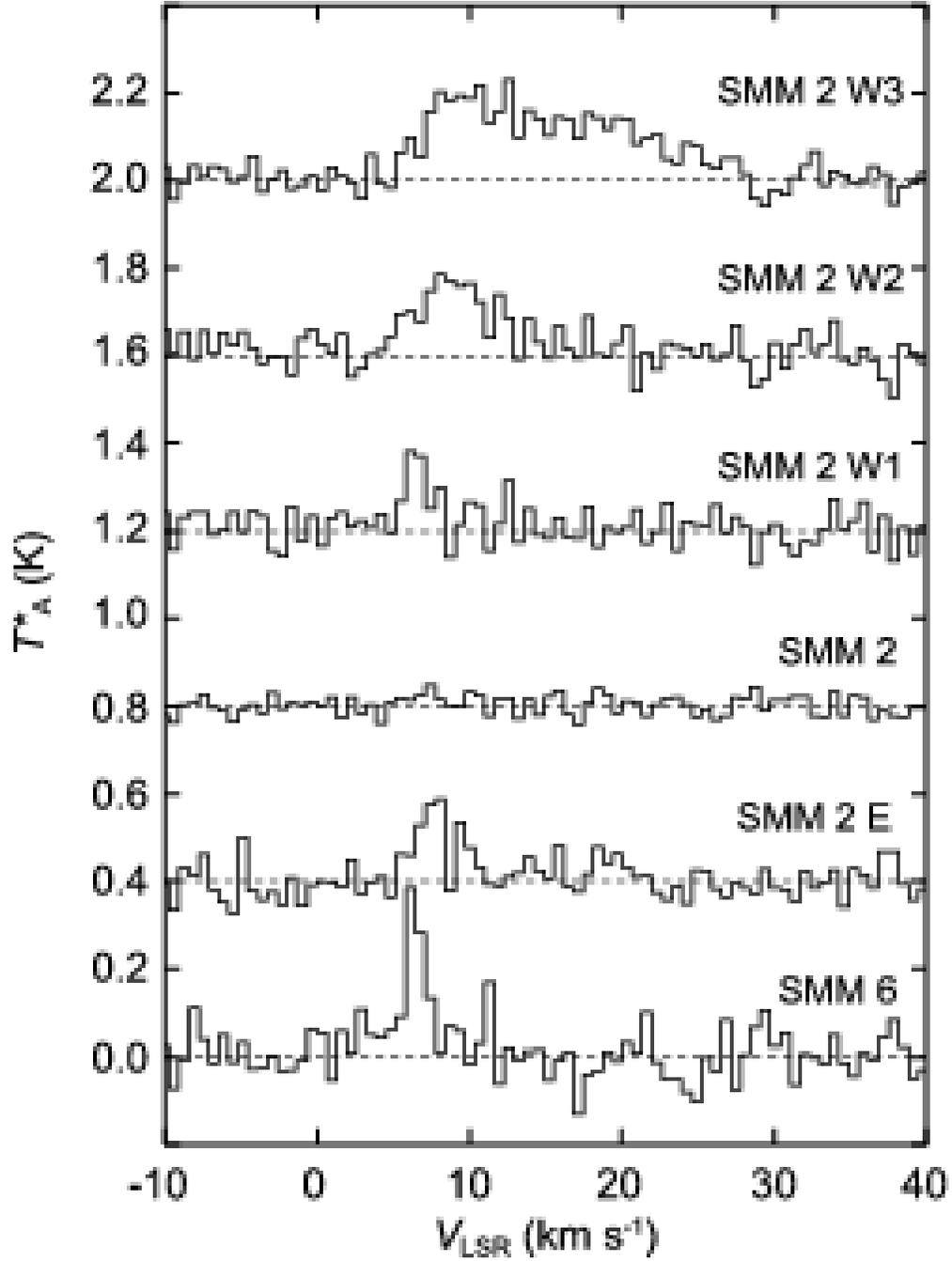}
\caption{SiO ($J=1-0$) line profiles along the outflow axis of SMM 2 and at the position of SMM 6. The velocity resolution 
is 0.5 km s$^{-1}$. 
\label{SiOlines}}
\end{figure}

The SiO ($J=1-0$) emission was detected at four positions out of 13 observed positions along the SMM 2 outflow  (Fig.\ref{SiOpos}).
The line profiles at the representative positions are shown in Fig.\ref{SiOlines}. In the northwestern lobe of the SMM 2 
outflow, the SiO  
emission is redshifted and its line width increases as the distance from SMM 2 increases (from SMM 2 W1 to SMM 2 W3, 
see Fig.\ref{SiOpos}). The terminal velocity of the SiO emission reaches up to 25 km s$^{-1}$ at the position of SMM 2 W3, which 
corresponds to the peak of the redshifted CO emission. In the southeastern lobe, the SiO emission was detected  
only at one position labeled SMM 2E. This position corresponds to the peak of the blueshifted CO emission. However, 
the SiO emission peaks at $V_{\rm LSR} = 7.7$ km s$^{-1}$, which is redshifted with respect to the cloud systemic velocity. 
At the position of SMM 2 itself, there was no detectable SiO emission.

The SiO emission originated from the SMM 6 outflow was searched for at four positions along the outflow, and the emission was 
detected only at the position of SMM 6. As shown in Fig.\ref{SiOlines}, the SiO line at SMM 6 shows the narrow component
($\Delta V = 1.3$ km s$^{-1}$) with its peak velocity at the systemic velocity of $V_{\rm LSR} = 6.4$ km s$^{-1}$, and 
a possible blueshifted component with the terminal velocity of $\sim 5$ km s$^{-1}$ with respect to the systemic velocity. 
\citet{Yamamoto1992} also reported the detection of the SiO $J=1-0$ and $2-1$ emission lines at $7\arcsec$ 
northwest from the position of SMM 6.
They pointed out that the line profile consists of a narrow ($\Delta v = 1.6$ km s$^{-1}$) component and broad 
($\Delta v = 6.4$ km s$^{-1}$) component. The redshifted wing of the broad component reported in \citet{Yamamoto1992} is 
weaker than the blueshifted wing,
hence the redshifted wing was not detected probably due to the low signal-to-noise ratio in our observations.

\section{DISCUSSION}
\subsection{CH$_3$OH and SiO Abundance}
\subsubsection{CH$_3$OH}
The column density values of the CH$_3$OH molecule at the positions of the protostars were calculated
by assuming the LTE condition using Equation (1). Since CH$_3$OH is not a linear molecule, the partition function was 
calculated from the energy level diagram \citep{Xu1997}. 
The excitation temperature was assumed to be 12 K, which is the same value as the kinetic temperature derived 
from NH$_3$ \citep{Bachiller1990}. We used the brightness temperature of the $2_{-1}-1_{-1}$ $E$ line in this calculation. 
Transitions between $A$ and $E$ species of CH$_3$OH are strictly prohibited, therefore the total column density of the 
CH$_3$OH molecule including both $A$ and $E$ species was estimated under the assumption that the abundance  
ratio between the $A$ and $E$ species is unity. The resulting
column densities toward five YSOs are summarized in Table \ref{METabundance}. 

The fractional abundance values of the CH$_3$OH molecule at the protostellar positions were estimated by comparing 
the derived CH$_3$OH column densities with the H$_2$ column densities at the same positions.  
The H$_2$ column densities were calculated using the 850 $\micron$ dust continuum data observed with the JCMT SCUBA 
by the COMPLETE project \citep{Ridge2006}. Under the assumption of optically thin dust emission, the H$_2$ column  
density $N({\rm H}_2)$ is calculated by
\begin{eqnarray}
	N({\rm H}_2) = {{S^{\rm beam}_{\nu}} \over \Omega _{\rm beam}\mu_{{\rm H_2}} m_{\rm H} \kappa_\nu B_\nu
	(T_{\rm D})},
	 \label{dust2N}
\end{eqnarray}
where $S^{\rm beam}_{\nu}$ is the flux density per beam, $\Omega _{\rm beam}$ is the beam solid angle, 
$\kappa_{\lambda}$ is the mass absorption coefficient per gram at 850 $\mu$m, and
$B_\nu (T_{\rm D})$ is the Planck function at a dust temperature $T_{\rm D}$. We assume the dust temperature of 12 K 
and $\kappa_{\lambda}=0.012 $ cm$^2$ g$^{-1}$. The SCUBA image was smoothed to the $18\arcsec$ resolution  
 in order to match the beam size of the CH$_3$OH observations. The 850 $\mu$m flux densities, the 
derived column densities and abundance are summarized in Table \ref{METabundance}. 

The derived abundance ranges from $2.3\times10^{-9}$ at SMM 1 to $9.4\times 10^{-9}$ at SMM 6, which are in the same order of
magnitude with the CH$_3$OH abundance in dark clouds, $2-5.1 \times 10^{-9}$ \citep{Friberg1988,Dickens2000}. The CH$_3$OH
abundance values toward SMM 6 and SMM 11 are a few times larger than those toward the other SMM sources. These two sources with 
high CH$_3$OH abundance are accompanied both with the CO outflow shown in the present observations and the shock-excited IR emission 
\citep{Walawender2005b}. On the other hand, there was no shocked IR emission found around SMM 2 and no significant outflow signature 
in our CO ($1-0$) data toward SMM 1 and SMM 3.
Therefore, the higher CH$_3$OH abundance values at SMM 6 and SMM 11 are considered to be influenced by the outflows.
Such a variation of the CH$_3$OH abundance is also found toward other YSOs. The CH$_3$OH abundance was derived toward 30 Class 0 
and Class I sources \citep{Buckle2002}, and the mean CH$_3$OH abundance was $1.9\times10^{-8}$ for the sources with broad component 
whose FWHM is larger than 2.0 km s$^{-1}$, and was $4.8 \times 10^{-9}$ for the sources without broad component. 
The CH$_3$OH abundance toward Class 0 sources is reported to be higher than that toward Class I sources \citep{Buckle2002},
however, our results show that the abundance depends on the presence of the broad line component rather than the evolutionary 
state of the sources.

Next, we examine the CH$_3$OH abundance in the outflow component around the position ``a''.
At the position ``a'', the integrated intensity of the redshifted CH$_3$OH wing emission ($V_{\rm LSR} > 9.6$ km s$^{-1}$) is 
1.1 K km s$^{-1}$. Under the assumption of the LTE condition
with an excitation temperature of 50 K, which is same as that of the CO 
emission, the column density of the CH$_3$OH molecule was estimated to be $6.6\times10^{14}$ cm$^{-2}$. Since dust continuum  
emission at the position ``a'' is too faint to estimate the H$_2$ column density, we used the CO line intensity in the same 
velocity range as the CH$_3$OH emission to calculate the H$_2$ column density of the high velocity gas at this position. Under 
the assumption of the optically thin CO emission with an excitation temperature of 50 K, the CO column density was derived to 
be $1.8 \times10^{16}$ cm$^{-2}$. The abundance ratio of CH$_3$OH to CO is [CH$_3$OH/CO]$=3.6\times10^{-2}$, and comparable 
to the abundance ratio measured in other outflows \citep{Garay2002}. Assuming the CO abundance with respect to H$_2$ is 
$10^{-4}$, the CH$_3$OH abundance ratio, [CH$_3$OH/H$_2$], is $3.6\times10^{-6}$. This value is more than 2 orders of magnitude 
larger than the abundance toward the protostars in B1 and the abundance in dark clouds. It is obvious that the CH$_3$OH abundance
is enhanced in the outflow.

\subsubsection{SiO}
\begin{figure}
\epsscale{.80}
\plotone{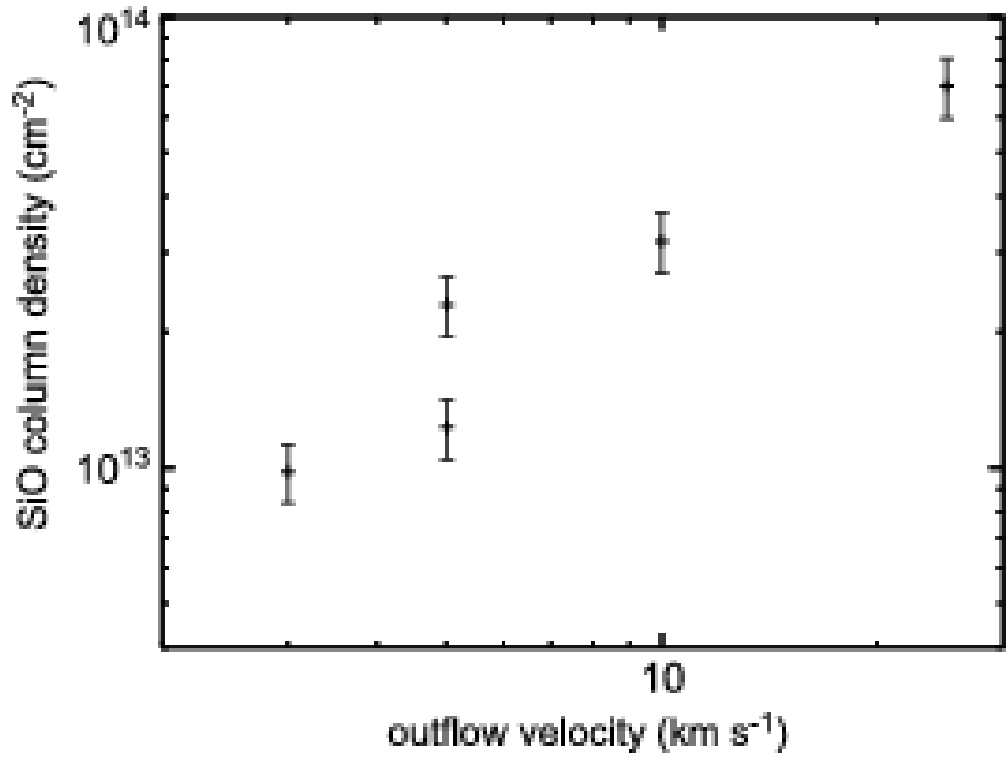}
\caption{Relation between the outflow terminal velocity measured from the systemic velocity at each point and the column 
density of the SiO molecule. The errors are calculated from the absolute intensity calibration uncertainty ($\sim 15\%$). 
\label{SiOcdv}}
\end{figure}

The column density of the SiO molecule was estimated using Equation (1) under the LTE assumption. The rotation 
constant of the SiO molecule is 21,787 MHz \citep{Lovas1974} and the dipole moment of the SiO molecule  is 3.1 D. We assumed 
the optically thin condition and adopted an excitation temperature of 50 K, the same assumption used to calculate 
the CO and CH$_3$OH column densities of the outflowing gas. Each SiO line profile was divided into the ``high-velocity'' and ``low 
velocity'' components in order 
to examine the difference of the abundance relative to CO in these two velocity ranges. The threshold of the two ranges was set to be 
$V_{\rm LSR}=9.6$ km s$^{-1}$ following the velocity criterion used to separate the CO line core and wing component. 
The derived SiO column densities at five positions (SMM 2 W1, W2, W3, E, and SMM 6) are summarized in Table \ref{SiOcd}.  
The column density of the SiO molecule was derived to be $1-2 \times 10^{13}$ cm$^{-2}$ in the low-velocity  
component. The high velocity component was detected only at the positions of SMM 2 W2 and SMM 2 W3. The SiO column density 
in the high-velocity component of SMM 2 W3 was $5.4\times10^{13}$ cm$^{-2}$, which is a few times higher than the values 
measured in the low-velocity component.
The relation between the total (high velocity + low velocity) SiO column density and the outflow velocity, which is determined from 
the terminal velocity of the SiO line, is shown in Fig. \ref{SiOcdv}. The SiO column density tends to increase as the outflow velocity 
increases. Similar correlation between the SiO column density and the outflow velocity was also found in other protostellar outflows 
studied by \citet{Garay2002}. These results support the idea that the more SiO molecules evaporate from dust grains into gas 
phase by collisions of the grains with higher velocity molecules in the outflows \citep{Schilke1997}. It should be noted that the SiO 
column density values reported in \citet{Garay2002} are 1 order of magnitude lower than those derived here. This is probably 
because the region covered by the telescope beam used by \citet{Garay2002} with HPBW of $57\arcsec$ was twice larger than that 
covered by our  $38\farcs4$ beam. Since the size of the SiO emitting region is supposed to be smaller than the beams of the single-dish 
telescopes, the beam-averaged column density becomes lower if the source is observed with the larger beam. In addition, most of 
the sources in the sample of  \citet{Garay2002} are located further away in distance.

The narrow (FWHM $\lesssim 1.0$ km s$^{-1}$) SiO line as seen in SMM 6 was also detected in NGC 1333 
\citep{Lefloch1998, Codella1999}, L1448-mm, and L1448 IRS3 \citep{Jimenez-serra2004}. These authors proposed different mechanism 
for the SiO enhancement. \citet{Lefloch1998} and \citet{Codella1999} suggested that the emission is arisen from the postshock 
equilibrium gas 
after the interaction of protostellar jets with surrounding dense clumps. On the other hand, \citet{Jimenez-serra2004} proposed 
that the emission is produced by the shock precursors associated with the protostellar jets. The narrow-line component around 
SMM 6 was detected only 
toward SMM 6 and does not have extended distribution like the one detected around NGC 1333 \citep{Lefloch1998}. The other 
difference is that the distribution of the narrow SiO line emission does not coincide with the high-velocity CO component in NGC 
1333. These differences indicate that the mechanism to produce the narrow SiO line emission is different in SMM 6 and NGC 1333 
and it is likely that the shock precursor component produces the narrow line detected toward SMM 6.

The abundance of SiO relative to CO was estimated by comparing the column densities of these molecules. 
Since the beam size of the SiO observations is larger than that of the CO, the CO map was smoothed in order to match 
the angular resolution with the SiO. However, our CO map is undersampled, and does not recover the entire flux. Therefore, we 
assumed that the spatial distribution of the CO flux between the observed points varies smoothly, and interpolated the CO flux values at 
the missing points using the neighboring data. This smoothing makes the derived CO column density $\sim 25$ -- 40 \% smaller 
than the value without smoothing. The CO spectra
in the ``low velocity'' range contain the emission from the quiescent ambient material that is unrelated with the outflows. Therefore 
in the calculation we adopted a method suggested by \citet{Margulis1985} in which the line intensity originated from the low-velocity 
outflow is assumed to be the same as the line intensity at the boundary velocity between the line core and wing components; 
$V_{\rm LSR}=2.1$ km s$^{-1}$ and 9.6 km s$^{-1}$ for the blue- and redshifted outflow components, respectively. The CO integrated
intensities in the low velocity range are estimated by assuming the uniform intensity over the integrated velocity range, $4.0 - 9.6$ 
km s$^{-1}$ toward the positions along the SMM 2 outflow and $2.1 - 9.6$ km s$^{-1}$ toward SMM 6. The
integrated velocity range is defined to cover all the SiO emissions at each position. The derived SiO abundance values are summarized
in Table \ref{SiOcd}. Assuming [CO/H$_2]=10^{-4}$, the SiO abundance relative to H$_2$ is $10^{-7} -10^{-8}$, 5--6 orders of 
magnitude higher than those measured in dark clouds \citep[$<$ a few $\times 10^{-12}$;][]{Ziurys1989}. The highest SiO/CO abundance 
of $4.7\times10^{-3}$ was obtained 
in the high velocity component at SMM 2 W3. This abundance is 3 orders of magnitude higher than the abundance estimated 
toward several outflows \citep{Garay2002} and comparable to the abundance at the extremely high-velocity (EHV) jet component in 
the L1448-mm outflow \citep[$\sim10^{-3}$;][]{Bachiller1991}. Such a high SiO abundance value and high terminal velocity
($\sim 50$ km s$^{-1}$, corrected for the inclination angle of 57\fdg3) suggest that the high-velocity SiO emission observed at 
SMM 2 W3 also arises from the jet component.

\subsection{Outflows as an Engine of Interstellar Turbulence}
Since protostellar outflows are considered to be the most likely mechanism to maintain the turbulent energy of the cloud 
\citep{Li2006}, we examine the turbulent energy budget of B1 main core following the method described in \citet{Stanke2007}.
The energy-loss rate due to the turbulence decay, $L_{\rm turb}$, and the energy input rate by the outflows, $L_{\rm gain}$, are 
compared.

According to numerical studies \citep{MacLow2004}, the supersonic turbulence dissipates in a short period comparable to the free-fall 
timescale $t_{\rm ff}$. Thus, $L_{\rm turb}$ is estimated by
\begin{eqnarray}
	L_{\rm turb} = E_{\rm turb} / t_{\rm ff},
	 \label{Lturb}
\end{eqnarray}
where $E_{\rm turb} = 3/2 M_{\rm core}\sigma_{\rm 1D}^2$ is the energy of the turbulence, $\sigma_{\rm 1D}$ is the one-dimensional
velocity dispersion. 
Since the observed region of the present study is well matched with the extent of B1 main core identified with 
the NH$_3$ observations \citep{Bachiller1990}, it is reasonable to adopt the mass $M_{\rm core}$ and the one-dimensional velocity 
dispersions $\sigma_{\rm1D}$ of B1 main core \citep[65 $M_{\sun}$ and 0.43 km s$^{-1}$;][]{Bachiller1990} in this calculation.
The 
free-fall time $t_{\rm ff}$ is calculated to be $3.5 \times 10^5$ yr using a diameter of 0.5 pc and a mass of 65 $M_{\sun}$. Therefore, 
the turbulence energy loss rate is calculated to be $L_{\rm turb} = 8.5 \times 10^{-3} L_{\sun}$.

On the other hand, $L_{\rm gain}$ is calculated under the assumption that all the outflow momentum is transferred to the 
turbulent motion of the cloud,
\begin{eqnarray}
	L_{\rm gain} &=& {1\over 2} \dot{M}_{\rm c}\sigma ^2_{\rm 3D},
	 \label{Lgain}
\end{eqnarray}
where $\dot{M}_{\rm c} = F_{\rm outflow} / \sigma_{\rm 3D}$ is the mass of the cloud which is accelerated up to the
velocity $\sigma_{\rm 3D}$, $F_{\rm outflow}$ is the total momentum input rate by outflows, and 
$\sigma_{\rm 3D}={\sqrt 3}\sigma_{\rm 1D}$ is the three-dimensional velocity dispersion of the clump. 
The total momentum flux provided by the three outflows from SMM 2, SMM 6, and SMM 11 is $F_{\rm outflow} = 1.8\times10^{-5} M_{\sun}$
km s$^{-1}{\rm yr}^{-1}$. Therefore, the energy input rate is calculated to be $L_{\rm gain}= 1.1\times10^{-3} L_{\sun}$.
This energy input rate is 1 order of magnitude smaller than the turbulence energy-loss rate.

There are some uncertainties in estimating the outflow energy input rate. The first issue is the effect of optical depth of the 
CO emission. Since our calculation assumed the optically thin CO wing emission, the energy input rate derived here is considered to 
be the lower limit. The mean opacity of the CO wing emission toward 16 YSOs is $\sim 4$ \citep{Cabrit1992}, and this value
makes the energy input rate four times larger than that derived under the optically thin assumption.
The second issue is the contribution of the low radial velocity component hidden in the line core. This effect becomes significant 
if the outflow axis is close to the plane of the sky, because most of the outflowing gas is moving along the transverse direction 
and shows the same radial velocity as the ambient gas. If the contribution of the "hidden" low radial velocity component is 
corrected by using the method described by \citet{Margulis1985}, which is also used to estimate the column density of the 
low-velocity outflow component in Section 4.1, the resulting energy input rate becomes twice as large as that calculated using the  
high-velocity wing component alone. The third issue is the contribution of the outflows that were not identified in the present  
observations. In addition to the outflows from SMM 2, SMM 6, and SMM 11, other two outflows from SMM 1 and SMM 3 were 
identified in the CO ($3-2$) observations by \citet{Hatchell2007b}. However, the contribution of these outflows to the total energy 
input is unlikely to be significant; the momentum flux value of the SMM 3 outflow is only 
$1.8\times 10^{-7} M_{\sun}$ km s$^{-1}{\rm yr}^{-1}$ \citep{Hatchell2007b}, which are more than a factor of 20 smaller than that 
of the SMM 2 outflow. In addition, the redshifted component of the SMM 1 outflow is treated as a part of the outflows from SMM 2 
and SMM11 in our calculations. 
In order to estimate the contribution of the weaker outflow components, we have integrated over all high-velocity 
component in the mapped area, and estimate the dynamical parameters of the high-velocity gas in this region. The total mass 
and momentum of the high velocity gas are derived to be $5.1\times10^{-2}$ $M_{\sun}$ and $0.48$ $M_{\sun}$ km s$^{-1}$, respectively.
If we assume that the dynamical timescale is $\sim 2\times10^4$ years, which is a typical timescale of the outflows in this region, the 
energy input rate is calculated to be $1.4\times10^{-3}$ $L_{\sun}$. The energy input rate derived here is only a factor of 1.3 larger than  
that from SMM 2, SMM 6, and SMM 11 outflows, suggesting that most of the mechanical energy is supplied by the three major outflows.
If we take the above issues into account, the outflow energy input rate can be 1 order of magnitude higher and becomes  
comparable to the turbulence energy decay rate. On the other hand, the efficiency of the energy transportation 
would be less than unity, because the extent of the outflow from SMM 2 and SMM 11 has already exceeded the dense part of the B1 main core region.
Therefore, we conclude that the outflow energy input is not enough to maintain the turbulence in B1 main core.
This result implies that the turbulence can not supply enough force against the gravity.

The magnetic field in B1 was measured by \citet{Crutcher1994} to be $30\pm4$ $\mu$G and $16\pm3$ $\mu$G in the NH$_3$ core
($r=1\farcm5$) and in the envelope ($r=9\arcmin$), respectively, using OH Zeeman effect. The magnetic field strength, the size and 
the mean density of the NH$_3$ core in B1 are well reproduced by the magnetically supercritical, collapsing core \citep{Crutcher1994}. 
This result indicates that the magnetic field is not able 
to support the core. Since the mass-to-flux ratio ($M/\Phi$) is independent of the distance to the object, the difference of the 
distances assumed in \citet{Crutcher1994} and the present study would not essentially affect the conclusion. 

Since B1 is located in the Perseus cloud complex which contains several active star-forming regions, it is interesting to compare 
the star formation in B1 with that in other star-forming regions.
\citet{Jorgensen2008} reported that there are eight protostars (i.e., Class 0 and Class I sources) out of nine YSOs in ``B1 tight 
group'', which corresponds to the B1 main core region. Since the fraction of protostars (8/9) in B1 is higher than the other star 
forming regions in Perseus, such as NGC 1333 (35/102) and IC 348 (11/121), they suggested that the star formation  
activity in B1 has recently initiated. In order to compare the turbulent energy budget in B1 with that of the more evolved star 
forming region, we have calculated the turbulence supply rate and decay rate of the NGC 1333 region. In the NGC 1333 region, 
six outflows were identified by the CO observations by \citet{Knee2000}. The total momentum injection rate from these outflows is 
$\sim 5.6\times 10^{-4}$ $M_{\sun}$ km s$^{-1}$ yr$^{-1}$, which is $\sim30$ times larger than that in B1. Using the velocity 
dispersion of 0.85 km s$^{-1}$, which is derived from the NH$_3$ observations \citep{Schwartz1978}, the energy input rate 
$L_{\rm gain}$ is estimated to be $6.8\times 10^{-2}$ $L_{\sun}$ in NGC 1333. On the other hand, if we adopt the mass of the core to be 
319 $M_{\sun}$, the turbulent decay rate $L_{\rm turb}$ is estimated to be $1.4\times 10^{-1}$ $L_{\sun}$. 
These two values agree within factor of 2, thus the outflows are 
powerful enough to maintain the turbulence in NGC 1333. Since
\citet{Knee2000} assumed the lower excitation temperature of 20 K and did not apply the inclination correction, the energy
input rate is underestimated compared to the case using our assumption of the excitation temperature of 50 K and the inclination 
angle of $57\fdg3$, in which the outflow momentum flux becomes a factor of 6 larger than the one originally derived by 
\citet{Knee2000}. This makes the outflow energy input superior to the turbulence energy loss.

The magnetohydrodynamical simulation \citep{Nakamura2007} shows 
that the cloud contraction proceeds in the initial phase of the cluster formation due to the decay of turbulence and decrease of 
the cloud momentum. This is consistent with the observed picture of B1 main core. 
The cloud reaches an equilibrium state in the later phase of the simulation because of the energy input by the outflows. As their 
simulations, when the outflows from the young Class 0 sources in B1 have enough evolved, it might be possible that the outflows
have a considerable contribution to the energy balances of the turbulence.

\section{SUMMARY}
In order to study how outflows from protostars influence the physical and chemical conditions of the parent molecular 
cloud, we have observed B1 main core, which harbors four Class 0 and three Class I sources, in the 
CO ($J=1-0$), CH$_3$OH ($J_K=2_K-1_K$), and the SiO ($J=1-0$) lines using the NRO 45 m telescope. We have identified three CO 
outflows in this region; one is an elongated ($\sim 0.3$ pc) bipolar outflow from a Class 0 protostar B1-c in the submillimeter clump
SMM 2, another is a rather compact ($\sim 0.1$ pc) outflow from a Class I protostar B1 IRS in the clump SMM 6, and the other is 
the extended outflow from SMM 11.
No significant outflows from other YSOs were identified in our CO ($J=1-0$) map. The CH$_3$OH emission lines are distributed 
around B1 main core with two peaks; one peak is located at the position of SMM 6, toward which the CH$_3$OH emission
in the line core velocity (5.5 $< V_{\rm LSR} <$ 7.5 km s$^{-1}$) is enhanced,  and the other is $120\arcsec$ west of SMM 2, at the 
head of the redshifted CO 
lobe of the SMM 2 outflow, toward which the CH$_3$OH line shows broad redshifted wing emission with a terminal velocity of  
13 km s$^{-1}$. In the redshifted lobe of the SMM 2 outflow, the SiO line also shows broad redshifted
wing whose terminal velocity reaches up to 25 km s$^{-1}$.

The CH$_3$OH and SiO abundances are significantly enhanced at the redshifted lobe of the SMM 2 outflow. 
It is likely that the shocks caused by the interaction between the outflow and ambient gas enhance the abundance of SiO and 
CH$_3$OH in the gas phase. Toward SMM 6, the CH$_3$OH and SiO lines are narrow and peak at the systemic velocity, 
however, the abundances of these molecules are enhanced compared with those in the quiescent dark clouds. Since the SiO 
emission around SMM 6 is localized at the position of the protostar, the origin of this narrow component is likely  
to be the shock precursor rather than the fossil outflow.

The turbulence energy decay rate and the outflow energy input rate are compared to investigate the energy budget in B1 main core.
It is found that the energy input rate, $1.1\times10^{-3} L_{\sun}$, is approximately 1 order of magnitude smaller than the turbulence decay 
rate of $8.5\times10^{-3} L_{\sun}$. Even though several uncertainties that increase the energy input rate are taken
into account, the outflows in B1 main core are not energetic enough to sustain the turbulence in this region.
Since the magnetic field is not strong enough to prevent the gravitational collapse of B1 main core \citep{Crutcher1994}, it is 
likely that B1 main core is gravitationally unstable. The high fraction of protostars among YSOs (8/9) in B1 main core  
suggests that the star formation activity in this region has been started very recently. In the more evolved star-forming region, 
NGC 1333, the cloud can be supported by the turbulence driven by the outflows. These results are consistent with the picture 
of the cloud evolution proposed by \citet{Nakamura2007}, in which molecular clouds contract due to the decay of turbulence in the 
early evolutionary phase, and become stable once the energy input from protostellar outflows
become sufficient enough to balance the energy loss.


%% Included in this acknowledgments section are examples of the
%% AASTeX hypertext markup commands. Use \url without the optional [HREF]
%% argument when you want to print the url directly in the text. Otherwise,
%% use either \url or \anchor, with the HREF as the first argument and the
%% text to be printed in the second.

\acknowledgments
We are grateful to the NRO staff for supporting the telescope operation. 
Nobeyama Radio Observatory is a branch of the National Astronomical Observatory of Japan, National Institutes of Natural Sciences.
This research has made use of the SIMBAD database, operated at CDS, Strasbourg, France.


%% copy editing.  Individual instruments can be provided in parentheses,
%% after the keyword, but they will not be verified.

%%Facilities: \facility{Nickel}, \facility{HST(STIS)}, \facility{CXO(ASIS)}.
{\it Facilities:} \facility{No:45m (BEARS)}




\clearpage

%%--------------------------------------------------------------------------------------------------

%%--------------------------------------------------------------------------------------------------

\begin{deluxetable}{lllcccl}
\tablewidth{0pt}
\tablecaption{Young Stellar Objects in B1}
\tablehead{
	\colhead{Name} & \colhead{$\alpha$} & \colhead{$\delta$} & \colhead{$L_{\rm bol}$} 
	 & \colhead{$T_{\rm bol}$} & \colhead{Class} & \colhead{other names} \\
	& \colhead{(J2000)} & \colhead{(J2000)} & \colhead{($L_{\sun}$)} & \colhead{(K)} & & }
\startdata 
	SMM 1 & 03 33 21.5 & 31 07 29 & $<$2.5 & $<$25 & 0 & B1-bN and B1-bS \\
	SMM 2 & 03 33 18.0 & 31 09 32 & 3.7 & 53 & 0 & B1-c \\
	SMM 3 & 03 33 16.3 & 31 06 53 & 2.6 & 32 & 0 & B1-d \\
	SMM 6 & 03 33 16.6 & 31 07 47 & 1.3 & 158 & I & B1-a, B1 IRS, IRAS 03301+3057 \\
	SMM 11 & 03 33 27.3 & 31 07 08 & 0.4 & 117 & I &  \\
	LkH$\alpha$ 327 & 03 33 30.4 & 31 10 51 & & & I & IRAS 03304+3100 \\
	\enddata
\label{sourceparam}
\tablecomments{Positions of the SMM sources are referred from \citet{Walawender2005b}, and that of LkH$\alpha$ 327
is obtained from Simbad. $L_{\rm bol}$, $T_{\rm bol}$, and Class of the SMM sources are referred from \citet{Hatchell2007a}.
The class of LkH$\alpha$ 327 is from \citet{Walawender2005b}.}
\end{deluxetable}

\begin{deluxetable}{llcc}
\tablecolumns{4}
\tablewidth{0pt}
\tablecaption{Observed Molecular Lines}
\tablehead{
	\colhead{Molecule} & \colhead{Transition} & \colhead{Frequency (GHz)} & \colhead{$E_{\rm u}$ (K)}}
\startdata 
	CO       & $J=1-0$                 & 115.2712 & 5.53 \\
	CH$_3$OH & $J_K=2_{-1}-1_{-1}$ $E$   & 96.73939 & 4.64 \\
	         & $J_K=2_{0}-1_{0}$ $A^+$ & 96.74142 & 6.79 \\
	         & $J_K=2_{0}-1_{0}$ $E$     & 96.74458 & 12.2 \\
	SiO      & $J=1-0$                 & 43.42386 & 2.08 \\
\enddata
\tablecomments{The upper state energies of the CH$_3$OH $E$ and $A^+$ symmetric state are the values relative 
to the ground state of $1_{-1}$ $E$ and $0_{0}$ $A^+$, respectively. The $1_{-1}$ $E$ level is 7.9 K above the $0_{0}$ $A^+$ level.}
\label{lineparam}
\end{deluxetable}

\begin{deluxetable}{lcccccc}
\tablecolumns{4}
\tablewidth{0pt}
\tabletypesize{\small}
\tablecaption{CO Line Intensity Ratio}
\tablehead{
	\colhead{Source} & \multicolumn{2}{c}{Position (CO $3-2$)} & \colhead{Integ. Vel. Range} & \colhead{$\int T_{\rm MB, 3-2}dv$} 
	& \colhead{$\int T_{\rm MB, 1-0}dv$} &\colhead{$T_{\rm ex}$}\\
	& \colhead{$\alpha$(J2000)} & \colhead{$\delta$(J2000)} & \colhead{(km s$^{-1}$)} &\colhead{(K km s$^{-1}$)}
	&\colhead{(K km s$^{-1}$)} & \colhead{(K)}}
\startdata 
	SMM 2 red & 03 33 15.866  &31 09 49.13 & 10 -- 20 & 24.8 & 4.5 & 46\\
	SMM 6 & 03 33 16.257 & 31 07 49.13 & $-10$ -- 2 & 33.2 & 6.2 & 42
\enddata
\label{COlineratio}
\end{deluxetable}

\begin{deluxetable}{lcccccc}
\tablecolumns{7}
\tablewidth{0pt}
\tablecaption{Outflow Parameters}
\tablehead{
	\colhead{Component} & \colhead{$M$} & \colhead{$P$}& \colhead{$\dot{M}$} & \colhead{$t_{\rm d}$}&
	 \colhead{$E_{\rm k}$} & \colhead{$F$}\\
	 & \colhead{($M_{\sun}$)} & \colhead{($M_{\sun}$km s$^{-1}$)} & \colhead{($M_{\sun}$ yr$^{-1}$)} & \colhead{(yr)} 
	 & \colhead{($M_{\sun}$km$^2$ s$^{-2}$)} & \colhead{($M_{\sun}$km s$^{-1}$ yr$^{-1}$)} }
\startdata 
      \multicolumn{7}{c}{SMM 2} \\
      \hline
	Blue        & 1.5$\times 10^{-2}$ & 0.14                      & 4.8$\times 10^{-8}$ & 2.9$\times 10^4$ & 0.66 & 4.8$\times 10^{-6}$ \\
	Red West & 5.3$\times 10^{-3}$ & 3.9$\times 10^{-2}$ & 1.8$\times 10^{-8}$ & 2.1$\times 10^4$ & 0.17 & 1.8$\times 10^{-6}$ \\
	Red East  & 5.2$\times 10^{-3}$ & 3.3$\times 10^{-2}$ & 2.1$\times 10^{-8}$ & 1.6$\times 10^4$ & 0.11 & 2.1$\times 10^{-6}$ \\
	Total       & 2.5$\times 10^{-2}$ & 0.21                      & 8.7$\times 10^{-8}$ &                         & 0.93  & 8.7$\times 10^{-6}$ \\
	\hline
	\multicolumn{7}{c}{SMM 6} \\
      \hline
	Blue & 1.2$\times 10^{-3}$  & 1.2$\times 10^{-2}$ & 1.1$\times 10^{-8}$ & 1.1$\times 10^4$ & 5.6$\times 10^{-2}$ & 1.1$\times 10^{-6}$ \\
	Red & 2.3$\times 10^{-3}$   & 1.7$\times 10^{-2}$ & 9.8$\times 10^{-9}$ & 1.8$\times 10^4$ & 7.0$\times 10^{-2}$ & 9.8$\times 10^{-7}$ \\
	Total & 3.5$\times 10^{-3}$ & 2.9$\times 10^{-2}$ & 2.0$\times 10^{-8}$ &                          & 0.13 & 2.0$\times 10^{-6}$ \\
	\hline
	\multicolumn{7}{c}{SMM 11} \\
      \hline
	Blue & 1.4$\times 10^{-2}$  & 0.14                     & 6.1$\times 10^{-8}$ & 2.3$\times 10^4$ & 0.72 & 6.1$\times 10^{-6}$ \\
	Red & 5.1$\times 10^{-3}$   & 3.0$\times 10^{-2}$ & 9.4$\times 10^{-9}$ & 3.2$\times 10^4$ & 8.7$\times 10^{-2}$ & 9.4$\times 10^{-7}$ \\
	Total & 1.9$\times 10^{-2}$ & 0.17                     & 7.0$\times 10^{-8}$ &                          & 0.81 & 7.0$\times 10^{-6}$ \\
	\enddata
\label{outflowparam}
\end{deluxetable}

\begin{deluxetable}{lcrcccc}
\tablecolumns{6}
\tablewidth{0pt}
\tablecaption{CH$_3$OH Column Density and Abundance}
\tablehead{
	\colhead{} & \colhead{$\int T_{\rm MB}dv$} & \colhead{$N$} & \colhead{$S_{850 \mu{\rm m}}$} & \colhead{$N$(H$_2$)} & \colhead{[CH$_3$OH/H$_2$]} & Shocked IR\\
	\colhead{Source} & \colhead{(K km s$^{-1}$)} & \colhead{(cm$^{-2}$)} & \colhead{(Jy beam$^{-1}$)} & \colhead{(cm$^{-2}$)} & \colhead{}} 
\startdata 
	SMM 1 & 1.9 & 2.3$\times 10^{14}$ & 1.0 & 1.0$\times 10^{23}$ & 2.3$\times 10^{-9}$ & No \\
	SMM 2 & 1.7 & 2.0$\times 10^{14}$ & 0.87 & 8.6$\times 10^{22}$ & 2.4$\times 10^{-9}$ & No \\
	SMM 3 & 1.3 & 1.6$\times 10^{14}$ & 0.64 & 6.3$\times 10^{22}$ & 2.5$\times 10^{-9}$ & Yes \\
	SMM 6 & 3.4 & 4.1$\times 10^{14}$ & 0.44 & 4.4$\times 10^{22}$ & 9.4$\times 10^{-9}$ & Yes \\
	SMM 11 & 0.58 & 7.1$\times 10^{13}$ & 0.12 & 1.1$\times 10^{22}$ & 6.2$\times 10^{-9}$ & Yes\\
\enddata
\label{METabundance}
\end{deluxetable}

\begin{deluxetable}{lcccccccc}
\rotate
\tablecolumns{9}
\tablewidth{0pt}
\tablecaption{SiO Column Density Along the Outflows}
\tablehead{
	\colhead{}& \colhead{} & \colhead{} & \multicolumn{3}{c}{High Velocity} & \multicolumn{3}{c}{Low Velocity} \\
	\colhead{} & \colhead{$\alpha$} & \colhead{$\delta$} & \colhead{$\int T_{\rm MB, SiO}dv$} & \colhead{$N$(SiO)} & \colhead{[SiO/CO]}
	& \colhead{$\int T_{\rm MB, SiO}dv$} & \colhead{$N$(SiO)} & \colhead{[SiO/CO]}\\
	\colhead{Source} &\colhead{(J2000)} & \colhead{(J2000)} & \colhead{(K km s$^{-1}$)} & \colhead{(cm$^{-2}$)}   
	& & \colhead{(K km s$^{-1}$)} & \colhead{(cm$^{-2}$)} & } 
\startdata 
	SMM 2 W3 & 3 33 09.3 & 31 09 57.9 & 2.6 & $5.4\times 10^{13}$ & $4.7\times10^{-3}$ & 0.78 & $1.6\times 10^{13}$ & $9.1\times10^{-4}$ \\
	SMM 2 W2 & 3 33 12.2 & 31 09 49.2 & 0.69 & $1.4\times 10^{13}$ & $1.4\times10^{-3}$ & 0.83&  $1.7\times 10^{13}$& $1.0\times10^{-3}$ \\
	SMM 2 W1 & 3 33 15.1 & 31 09 40.6 & \nodata & \nodata & \nodata & 0.47 & $9.8\times 10^{12}$ & $5.1\times10^{-4}$ \\
	SMM 2 E & 3 33 29.7 & 31 08 57.4 & \nodata & \nodata & \nodata & 0.59& $1.2\times 10^{13}$ & $2.7\times10^{-4}$ \\
	SMM 6 & 3 33 16.6 & 31 07 46.8 & \nodata & \nodata &\nodata & 1.1 & $2.3\times 10^{13}$ & $2.1\times10^{-3}$ \\
\enddata
\label{SiOcd}
\end{deluxetable}

\end{document}